\documentclass[singlecolumn,showpacs,superscriptaddress,notitlepage]{revtex4-1}
\usepackage{amsmath,amssymb,amsfonts,graphics,graphicx,dcolumn,bm}
\usepackage[toc,page]{appendix}
\usepackage{natbib}
\usepackage{geometry}
\usepackage{qtree}
\usepackage{chngcntr}

\newcolumntype{C}[1]{>{\centering\let\newline\\\arraybackslash\hspace{0pt}}m{#1}}

\usepackage[ddmmyyyy,hhmmss]{datetime}

\newcommand{\scl}
{\affiliation{Scientific Computing Laboratory, Center for the Study of Complex Systems, Institute of Physics Belgrade, University of Belgrade, Pregrevica 118, 11080 Belgrade, Serbia}}

\newcommand{\etf}
{\affiliation{School of Electrical Engineering, University of Belgrade, P.O. Box 35-54, 11120 Belgrade, Serbia}}

\begin{document}

\title{Associative nature of event participation dynamics: a network theory approach}
\author{Jelena Smiljani\'c}
\email[Email:]{jelenas@ipb.ac.rs}
\scl
\etf
\author{Marija Mitrovi\'c Dankulov}
\scl

\begin{abstract}
The affiliation with various social groups can be a critical factor when it comes to quality of life of each individual, making such groups an essential element of every society. The group dynamics, longevity and effectiveness strongly depend on group's ability to attract new members and keep them engaged in group activities. It was shown that high heterogeneity of scientist's engagement in conference activities of the specific scientific community depends on the balance between the numbers of previous attendances and non-attendances and is directly related to scientist's association with that community. Here we show that the same holds for leisure groups of the Meetup website and further quantify individual members' association with the group. We examine how structure of personal social networks is evolving with the event attendance. Our results show that member's increasing engagement in the group activities is primarily associated with the strengthening of already existing ties and increase in the bonding social capital. We also show that Meetup social networks mostly grow trough big events, while small events contribute to the groups cohesiveness.
\end{abstract}

\maketitle

\section*{Introduction}
One of the consequences of the rapid development of the Internet and growing presence of information communication technologies is that a large part of an individual's daily activities, both off and online, is regularly recorded and stored. This newly available data granted us a substantial insight into activities of a large number of individuals during long period of time and led to the development of new methods and tools, which enable better understanding of the dynamics of social groups \cite{RevModPhys.81.591}. 
The structure and features of social connections both have strong influence and depend on social processes such as cooperation \cite{Nowak1560, Fowler23032010}, diffusion of innovations \cite{doi:10.1086/225469, RevModPhys.87.925}, and collective knowledge building \cite{mitrovic2015}. 
Therefore, it is not surprising that the theory of complex networks has proven to be very successful in uncovering mechanisms governing the behavior of individuals and social groups \cite{boccaletti2006, holme2012}.\\  
The human activity patterns, the structure of social networks, and the emergence of collective behavior in different online communities have been extensively studied in the last decade \cite{Aral337,Gonzalez-Bailon01072013,Lin:2008:FFA:1367497.1367590,mitrovic2011,garas2012,torok2013,yasseri2012,mitrovic2015}. On the other hand, the dynamics of offline social groups, where the activities take place trough offline meetings (events), have drawn relatively little attention, given their importance. Such offline groups, both professional and leisure ones, provide significant benefits and influence everyday lives of individuals, their broader communities, and the society in general: they offer social support to vulnerable individuals \cite{montazeri2001,davison2000}, can be used for political campaigns and movements \cite{tamcho2012,weinberg2006}, or can have an important role in career development \cite{smiljanic2016}. As they have different purpose, they also vary in the structure of participants, dynamics of meetings, and organisation. Some groups, such as cancer support groups or scientific conference communities, are intended for a narrow circle of people while others, leisure groups for instance, bring together people of all professions and ages. In the pre-Internet era these groups have been, by their organisation and means of communication between their members, strictly offline, while today we are witnessing the appearance of a growing number of hybrid groups, which combine both online and offline communication \cite{weinberg2006}. Although inherently different, all these social groups have two main characteristics in common: they do not have formal organization, although their members follow certain written and non-written rules, and their membership is on a voluntary basis. Bearing this in mind, it is clear that the function, dynamics and longevity of such self-organized communities depend primarily on their ability to attract new and retain old active members. Understanding the reasons and uncovering key factors that influence members to remain active in the social group dynamics are thus important, especially having in mind their relevance for the broader social communities and the society. \\   
The size of social groups and personal social networks, as well as their structure, have been extensively studied. The considerable body of evidence \cite{dunbar1993, dunbar2008, hill2003, dunbar2011} suggests that the typical size of natural human communities is approximately $150$, that both groups and personal social networks are highly structured, and consist of social layers characterized by different strengths of relationships. The relationships within each layer are characterized by a similar mean frequency of interaction and emotional closeness, both of which decrease rapidly as we move trough network layer. These findings have been explained using the Social Brain Hypothesis, which relates the average size of species' personal network with the computational capacity of its brain. Here we confirm that these findings also hold for leisure groups where the frequency of interactions among members is constrained by the  event dynamics. We also explore how the number of attended events is related to the size and layered structure of member's personal network.\\
Previous research on hybrid social groups and interplay between offline and online interactions has shown that offline meetings enhance attendees' engagement with online community and contributes to the creation of a bonding capital \cite{doi:10.1080/13691180903468954,hristova2013}. In our previous work \cite{smiljanic2016} it was shown that the participation patterns of scientist in a particular conference series are not random and that they exhibit a universal behaviour independent of conference subject, size or location. Using the empirical analysis and theoretical modeling we have shown that the conference attendance depends on the balance between the numbers of previous attendances and non-attendances and argued that this is driven by scientist's association with the conference community, i.e. with the number and strength of social ties with other members of the conference community. We also argued that similar behaviour can be expected in other social communities when it comes to members' participation patterns in organised group events. Here we provide empirical evidence supporting these claims and further investigate the relationship between the dynamics of participation of individuals in social group activities and the structure of their social networks.\\            
The Meetup portal, whose group dynamics we study here, is an event-based social network. Meetup members use the online communication for the organization of offline gatherings. The online availability of event attendance lists and group membership lists enables us to examine the event participation dynamics of Meetup groups and its influence on the structure of social networks between group members. The diversity of Meetup groups in terms of the type of activity and size allows us to further examine and confirm the universality of member's participation patterns. We note that previous papers using Meetup source of data have mostly focused on the event recommendation problem 
\cite{Qiao:2014:ERE:2892753.2893014, Zhang:2013:CLF:2487575.2487646, 7113315, Macedo:2015:CER:2792838.2800187, EBSN}, structural properties of social networks, and relationships between event participants \cite{EBSN, ICWSM1613122} by disregarding evolutionary behaviour of Meetup groups.\\
In this paper, we examine the event-induced evolution of social networks for four large Meetup groups from different categories. Similarly to the case of conference participation, we study the probability distribution of a total number of meetup attendances and show that it also exhibits a truncated power law for all four groups. This finding suggests that the event participation dynamics of Meetup groups is characterized by a positive feedback mechanism, which is of social origin and is directly related to member's association with social community of the specific Meetup group. Using the theory of complex networks we examine in more detail the correlation between an individual's decisions to participate in an event and her association with other members of that Meetup group. Specifically, we track how member's connectedness with the community changes with the number of attendances by measuring change in the clustering coefficient and relation between the degree and the strength in an evolving weighted social network, where only statistically significant connections are considered. Our results indicate that greater involvement in group activities is more associated with the strengthening of existing than to creation of new ties. This is consistent with previous research on Meetup which has shown that repeated event attendance leads to an increase in bonding and a decrease in bridging social capital \cite{doi:10.1080/13691180903468954, mccully2011}. Furthermore, in view of the fact that people interact and networks evolve through events, we examine how particular a event affects the network size and its structure. We investigate effects of event sizes and time ordering on social network organization by studying changes in the network topology, numbers of distinctive links and clustering, caused by the removal of a specific event. We find that large events facilitate new connections, while during the small events already acquainted members strengthen their interpersonal ties. Similar behavior was observed at the level of communities, where small communities are typically closed for new members, while contrary to this, changes in the membership in large communities are looked at favorably \cite{palla2007quantifying, Backstrom:2006:GFL:1150402.1150412}.\\   
This paper is organized as follows: we first study the distribution of the total number of participations in four Meetup groups from different categories. 
Next we introduce a filtered weighted social network to characterize significant social connections between members and discuss its structural properties. Specifically, we study 
how the local topological properties evolve with the growth of the number of participations in order to derive relationships between members' association with the group and 
their activity patterns. In order to analyze impact of a particular event on the network organization, we remove events using different strategies and study how this influences the social structure.

\section*{Results}
\subsection*{Event participation patterns of Meetup groups}
\begin{figure}
\includegraphics[scale=1]{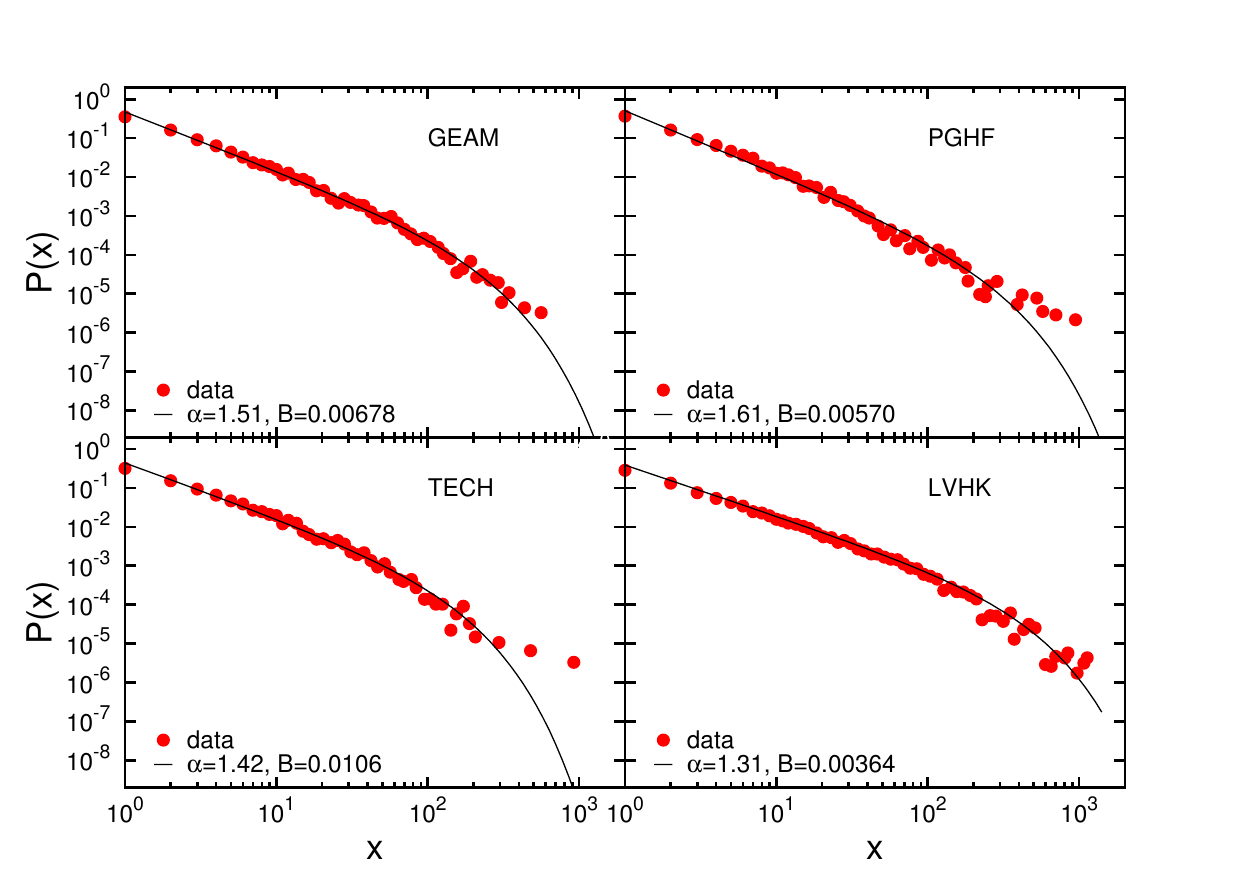}
\caption{\textbf{Total number of attended events.} Probability distributions $P(x)$ of total number of participations $x$, for four Meetup groups. Solid line represents best fit to truncated power law distribution, $x^{-\alpha}e^{-Bx}$.}
\label{total}
\end{figure}

Meetup is an online social networking platform that enables people with a common interest to start a group with a purpose of arranging offline meetings (events, meetups) all over the world. The groups have various topics and are sorted into $33$ different categories, such as careers, hobbies, socializing, health, etc. These groups are of various sizes, have different event dynamics, and hierarchical organisation. They also differ in the type of activity members engage, ranging from socializing events (parties and clubbing) to professional trainings (seminars and lectures). Common to all groups is the way they organize offline events: each member of the group gets an invitation to event to which they reply with yes or no, creating in that way a record of attendance for each event. We use this information to analyze event participation patterns and to study the evolution of the social network.\\ 
Here, in particular we analyze four large groups, each belonging to a different category and having more than three thousand organized events (see Methods and Table~\ref{dataset}). We chose these four groups because of their convenience for statistical analysis, large number of members and organized events, and also for the fact that they are quite different concerning the type of activities and interests their members share. The \textit{geamclt} (GEAM) group is made of \textit{foodie thrill-seekers} who mostly meet in restaurants and bars in order to try out new exciting foods and drinks, while people in the \textit{VegasHiker} (LVHK) group are hikers who seek excitement trough physical activity. The \textit{Pittsburgh-free} (PGHF) is our third group which invites its members to free, or almost free, social events, and the fourth considered group \textit{TechLife Columbus} (TECH), which is about social events and focuses on technology-related community networking, entrepreneurship, environmental sustainability, and professional development.\\
Figure \ref{total} demonstrates that the probability distributions of total attendance numbers of members in events for all four groups exhibits a truncated power law behavior (see Fig \ref{fig_A} and Table \ref{table_A} in SI, which show a comparison with the exponential and power law fit), with power law exponent larger than one. Similarly to the conference data \cite{smiljanic2016}, the exponential cut-off is a finite size effect. Power law and truncated power law behavior of probability distributions can be observed for the number of and the time lag between two successive participations in group-organized events, Fig \ref{fig_B} and Fig \ref{fig_C} in SI. In fact, we find that similar participation patterns which differ in values of exponents) can be observed for all Meetup groups, regardless of their size, number of events or category. As in the case of the conference participation dynamics \cite{smiljanic2016}, this indicates that the probability to participate in the next event depends exclusively on the balance of numbers of previous participations and non-participations. We argued in \cite{smiljanic2016} that the forces behind conference participation dynamics are of social origin, and it follows from Fig \ref{total} that the same can be argued for the case of the Meetup group participation dynamics. The more participations in group activities member has, the stronger and more numerous are her connections to the other group members, and thus her association with the community. We further explore this assumption by investigating the event-driven evolution of social networks of the four different Meetup groups.\\

\subsection*{Structure of social event-based network}
\begin{figure}
\includegraphics[scale=1]{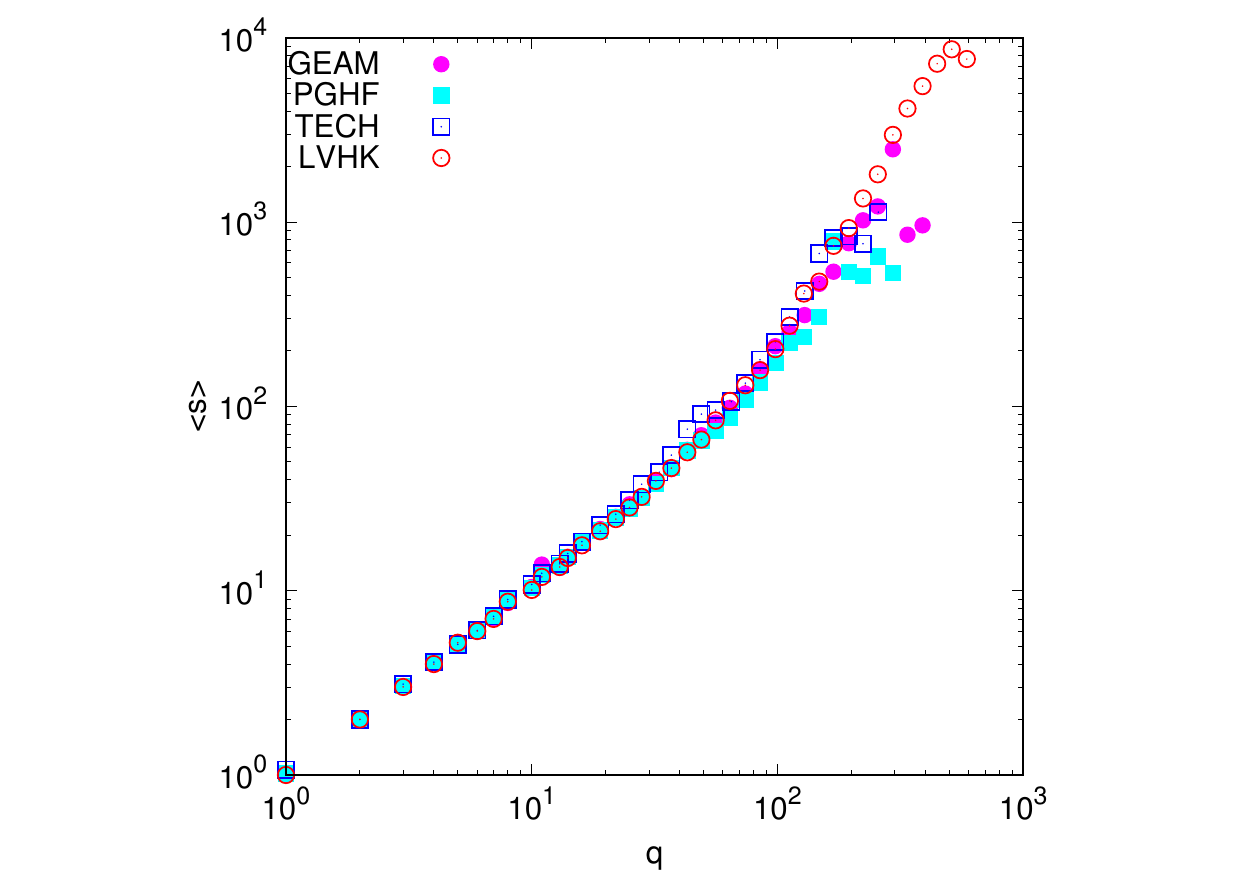}
\caption{ \textbf{Node strength dependence on node degree.} Dependence of average member's strength $\langle s\rangle$ on her degree $q$ in social network of significant links for considered groups.}
\label{deg_str}
\end{figure}

We construct a social network between group members for each considered group, as a network of co-occurrence on the same event (see Methods for more details). By definition, these networks are weighted networks with link weights between two members equal to the number of events they participated together. These networks are very dense, as a direct result of the construction method, with broad distribution of link weights (see Fig~\ref{fig_D} in SI). However, co-occurrence at the same event does not necessarily imply a relationship between two group members. For instance, a member of a group that attends many events, or big events, has a large number of acquaintances, and thus large number of social connections, which are not of equal importance regarding her association with the community. Similarly, two members that attend a large number of events can have relatively large number of co-occurrences, which can be the result of coincidences and not an indicator of their strong relationship. In order to filter out these less important connections we use a filtering technique based on the configuration model of bipartite networks \cite{Dianati:2197182,2016arXiv160702481S} (see Methods). By applying this technique to weighted networks we reduce their density and put more emphasis on the links that are less likely to be the result of coincidences. In this way we emphasize the links of higher weights without the removal of all links below certain threshold (see Fig~\ref{fig_D} in SI), a standard procedure for network pruning. We explore the evolution of social networks of significant relationships between Meetup group members by studying how the local characteristics of the nodes (members) change with their growing number of participations in group activities.\\
Association with the community of a specific Meetup group can be quantitatively expressed trough several local and global topological measures of weighted networks. Specifically, here we explore how the number of significant connections (member's degree) and their strength (member's strength), as well as how member's embeddedness in a group non-weighted and weighted clustering coefficient) are changing with the number of attended group events. Figure \ref{deg_str} shows how average strength of a node depends on its degree in filtered networks of four selected Meetup groups. While member's degree equal to the number of member's significant social relationships, the strength measures how strongly she is connected to the rest of the group \cite{Barrat16032004}. In all considered Meetup groups members with small and medium number of acquaintances ($q\leq 50$) have similar values of strengths and degrees, i.e., their association with the community is quantified by the number of people they know and not through the strength of their connections (see Fig \ref{deg_str}). Having in mind that the average size of an event in these four groups is less than $20$, we can conclude that majority of members with a degree less than $50$ are the ones that attended only a few group meetups. A previous study \cite{Macedo:2015:CER:2792838.2800187} has found that the probability for a member to attend a group event strongly depends on whether her friends will also attend. The non-linear relationship between the degree and the average strength for $q > 50$ shows that event participation of already engaged members (ones who already attended few meetings) is more linked to the strength of social relations than to their number. This means that at the beginning of their engagement in group activities, when the association is relatively small, the participation is conditioned by a number of members a person knows, while later, when the association becomes stronger, the intensity of relations with already known members becomes more important.\\ 

\begin{figure}
\includegraphics[scale=1]{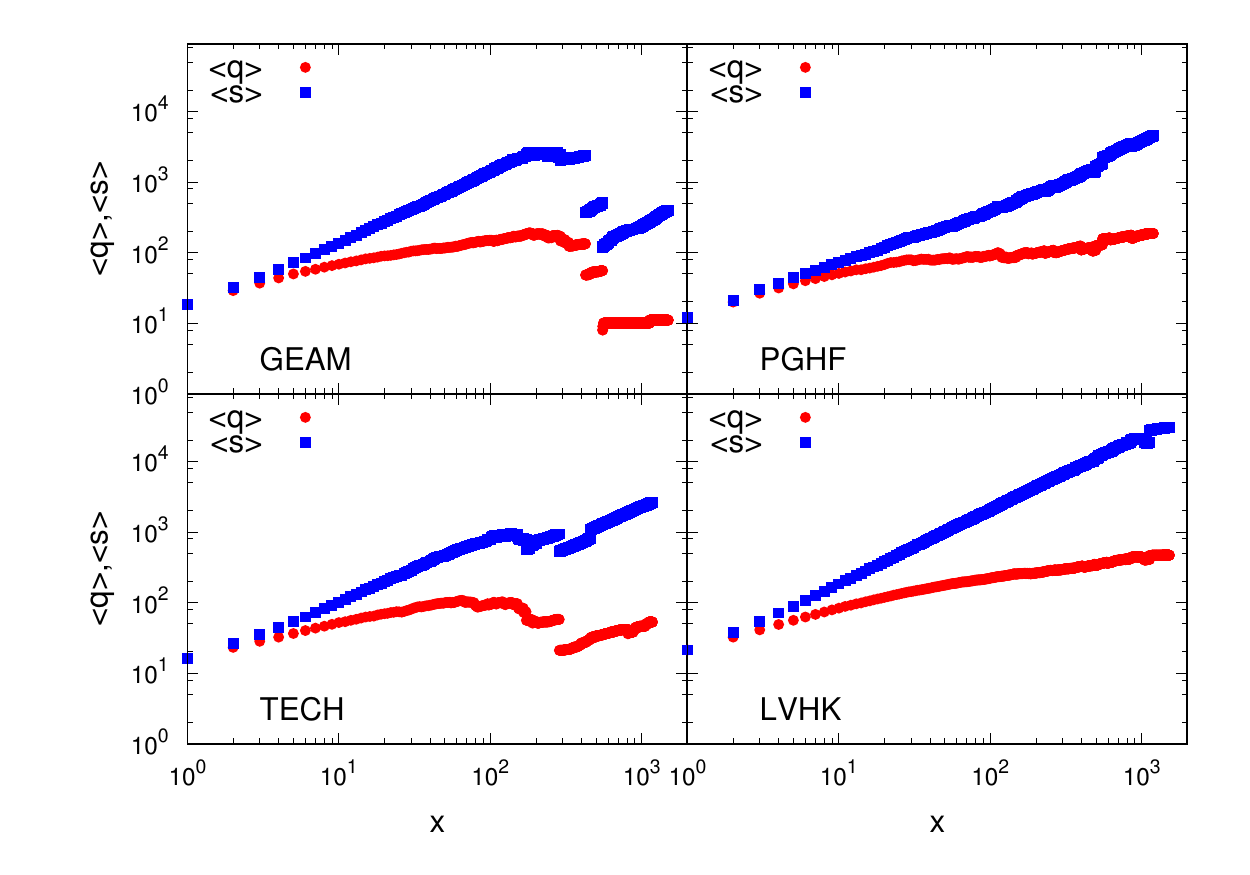}
\caption{\textbf{Event driven evolution of member's degree and strength.} Dependence of member's average degree $\langle q \rangle$ and strength $\langle s \rangle$ on number of attended group events by member $x$ for four considered Meetup groups.}
\label{deg_str_part}
\end{figure}

\begin{figure}
\includegraphics[scale=1]{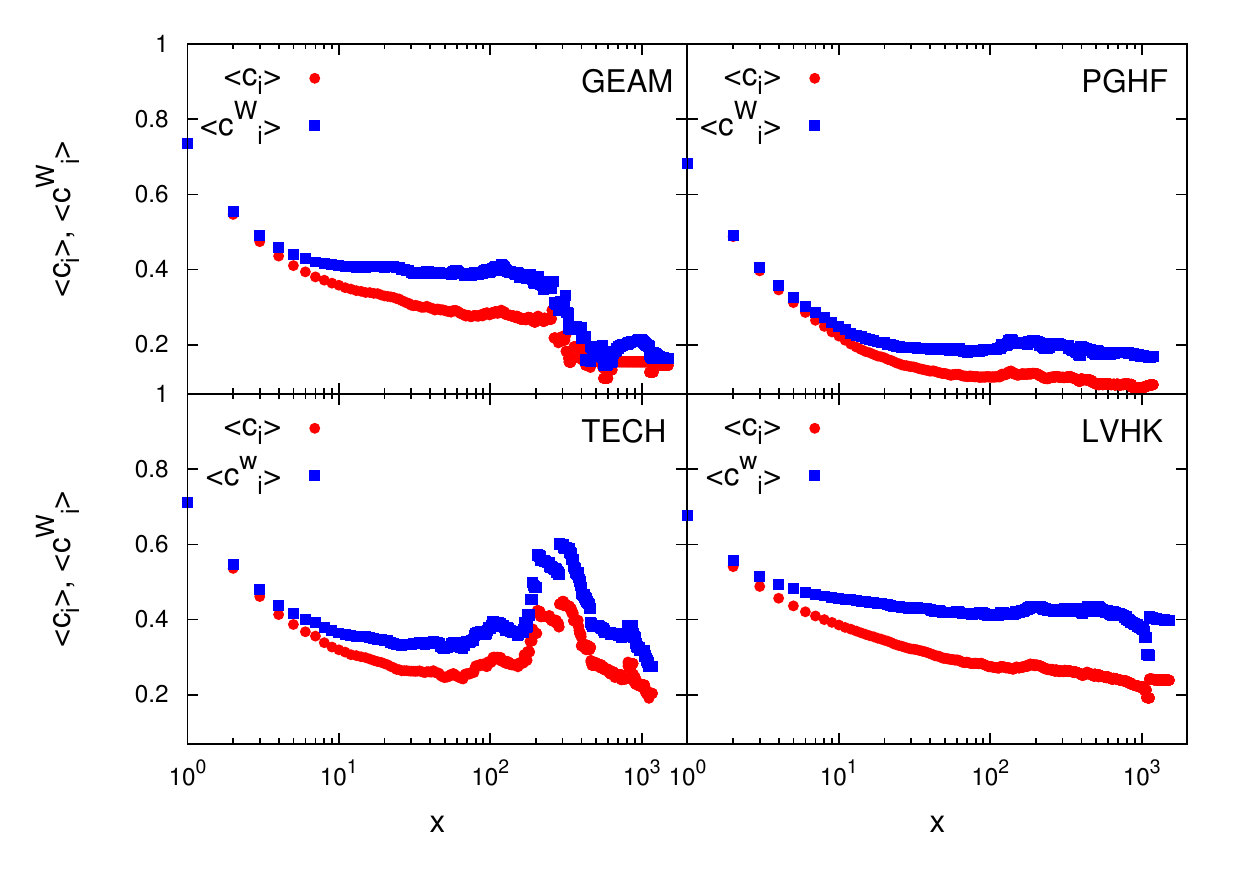}
\caption{\textbf{Local cohesiveness of social networks of significant links.} Evolution of local cohesiveness of members personal networks, measured by averaged non-weighted $\langle c_{i} \rangle$ and weighted clustering coefficients $\langle c^{W}_{i} \rangle$, with the number of events attended by the member $x$.}
\label{C_Cw_part}
\end{figure}

This finding is further supported if we consider the change of the average degree and strength with the number of participations. Figure \ref{deg_str_part} shows how the average member's degree and strength evolve with the number of participations in group's events. At the beginning, the degree and strength have the same value and grow at the same rate, but after only few participations the strength becomes larger than the degree, and starts to grow much faster for members of all four Meetup communities. After $100$ attended events the average strength of a member is up to ten times larger than her degree (see Fig~\ref{fig_E} in SI). This indicates that the event participation dynamics is mostly governed by the need of a member to maintain and strengthen her relationships with already known members of a community. As a matter of fact, our analysis of member's embeddedness in these social networks shows that members maintain strong relations with single members of the community, but also with small subgroups of members. A comparison with randomized data (Figs~\ref{fig_E} and \ref{fig_F} in SI) reveals that both the degree and strength grow slower with the number of events, and that their ratio is higher than in the original data. Relatively high value of the average clustering coefficient $\langle c_{i} \rangle$, shown in Fig \ref{C_Cw_part} indicates that there is a high probability (more than $10\%$ on average) that friends of a member also form significant relationships. The slow decay of $\langle c_{i} \rangle$ with the number of participations and the fact that it remains relatively large (above $0.2$) even for participants with a thousand of attended meetups, Fig \ref{C_Cw_part}, show that personal networks of members have tendency to remain clustered, i.e., have relatively high number of closed triplets compared to random networks.\\
We now further examine the structure of these triplets and its change with the number of participations by calculating the averaged weighted clustering coefficient. The weighted clustering coefficient $c^{W}_{i}$ measures the local cohesiveness of personal networks by taking into account the intensity of interactions between local triplets \cite{Barrat16032004}. This measure does not just take into account a number of closed triplets of a node $i$ but also their total relative weight with respect to the total strength of the nodes (see Methods). We also examine how the value of weighted clustering coefficient, averaged over all participants that have attended $x$ events, designated as $\langle c_i^W(x)$, with the number of attended events. As it is shown in Fig \ref{C_Cw_part}, a member's network of personal contacts shows high level of cohesiveness, on the average. Like its non-weighted counterpart, the value of $\langle c^{W}_{i} \rangle$ only slightly decreases during member's early involvement in group activities, while later it remains constant and independent of the number of participations. A comparison of the values of weighted and non-weighted clustering coefficients reveals the role of strong relationships in local networks, i.e., whether they form triplets or bridges between different cohesive groups \cite{Barrat16032004}. At the beginning of member's involvement in a group, these two coefficients have similar values, Fig \ref{C_Cw_part}, which indicates that the cohesiveness of a subgroup of personal contacts is not that important for the early participation dynamics. As a number of attended events grows, as well as a number and strength of personal contacts, the weighted clustering coefficient becomes larger than its non-weighted counterpart, indicating member's strongest ties with other members who are also friends. The fact that in later engagement the weighted clustering coefficient is larger than its non-weighted counterpart indicates that the clustering has an important role in the network organization of Meetup groups and thus in the group participation dynamics \cite{Barrat16032004}. Low and very similar values of the clustering and weighted clustering coefficients in networks obtained for randomized data (Fig~\ref{fig_G} in SI) further confirm our conclusion about the importance of clustering in the event participation dynamics. The observed discontinuity and decrease of values of the degree, strength and both clustering coefficients, Figs~\ref{deg_str_part} and \ref{C_Cw_part}, for groups GEAM and TECH are consequences of a small number of members who attended more than $300$ events.

\subsection*{Event importance in group participation dynamics}

\begin{figure}
\includegraphics[scale=0.5]{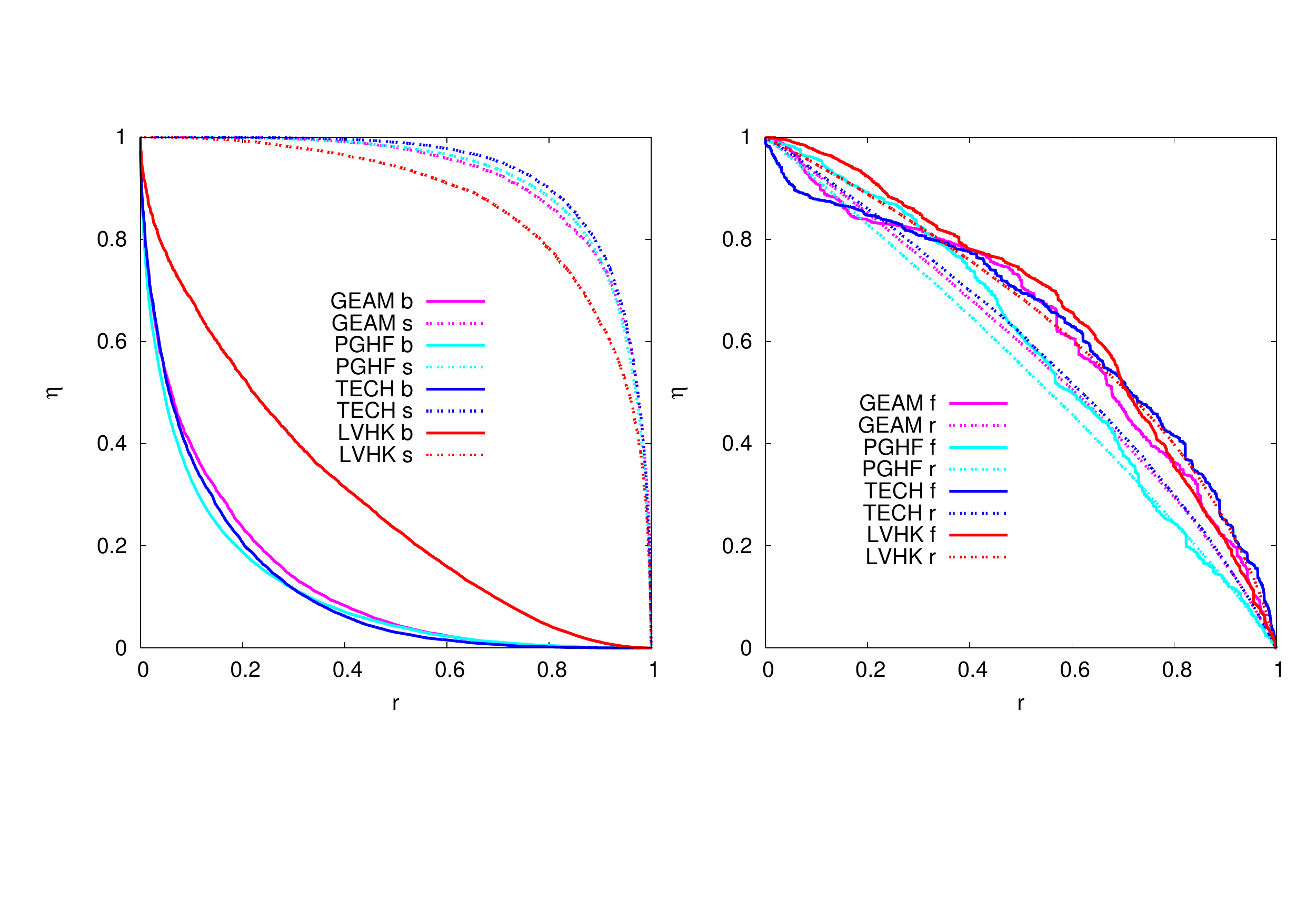}
\caption{\textbf{Importance of event size for number of distinctive links in the networks.} Change of $\eta$ with removal of events according to their size (left) and temporal and random order (right). Abbreviations indicate order in which we remove events: {\bf b} - from the largest to the smallest, {\bf s} - from the smallest to the largest, {\bf f} - from the first to the last and {\bf r} - random.}
\label{eta_event}
\end{figure}

In our previous work \cite{smiljanic2016}, we have shown that the conference participation dynamics is independent of the conference topic, type and size. The same holds true for the Meetup participation dynamics, i.e., the member's participation patterns in the Meetup group activities do not depend on the group size, category, location or type of activity. However, the size of group events and their time order may influence the structure of network and thus group dynamics. We explore how topological properties of networks, specifically the number of acquaintances and network cohesion, change after the removal of events according to a certain order (see Methods for details).\\
Firstly, we study how the removal of events according to a certain order influences the number of overall acquaintances in the network. For this purpose we define a measure $\eta$ (see Methods), which we use to quantify the percentage of the remaining significant acquaintances after the removal of an event. Figure \ref{eta_event} shows the change of measure $\eta$ after the removal of a fraction $r$ of events according to a chosen strategy. We see that most of new significant connections are usually made during the largest events. The importance of large events for the creation of new acquaintances is especially striking for the groups GEAM, PGHF, and TECH, where about $80\%$ of acquaintances only met at top $20\%$ of events. For LVHK the decrease is slower, probably due to a difference in the event size fluctuations (see Fig~\ref{fig_H} in SI), but still more than $50\%$ of acquaintances disappear if we remove top $40\%$ of events, which is still much higher percentage of contacts compared to random removal of events (see Fig \ref{eta_event} (right)). Similar results are observed when we remove events in the opposite order, Fig \ref{eta_event} (left). Only $20\%$ of acquaintances are being destroyed after the removal of $80\%$ of events, for all four groups. This indicates that new and weak connections are usually formed during large events, while these acquaintances are further strengthen during small meetups. On the other hand, the removal of events according to their temporal order, Fig \ref{eta_event}, has very similar effect as random removal, i.e., the value of parameter $\eta$ decreases gradually as we remove events. 

\begin{figure}
\includegraphics[scale=0.5]{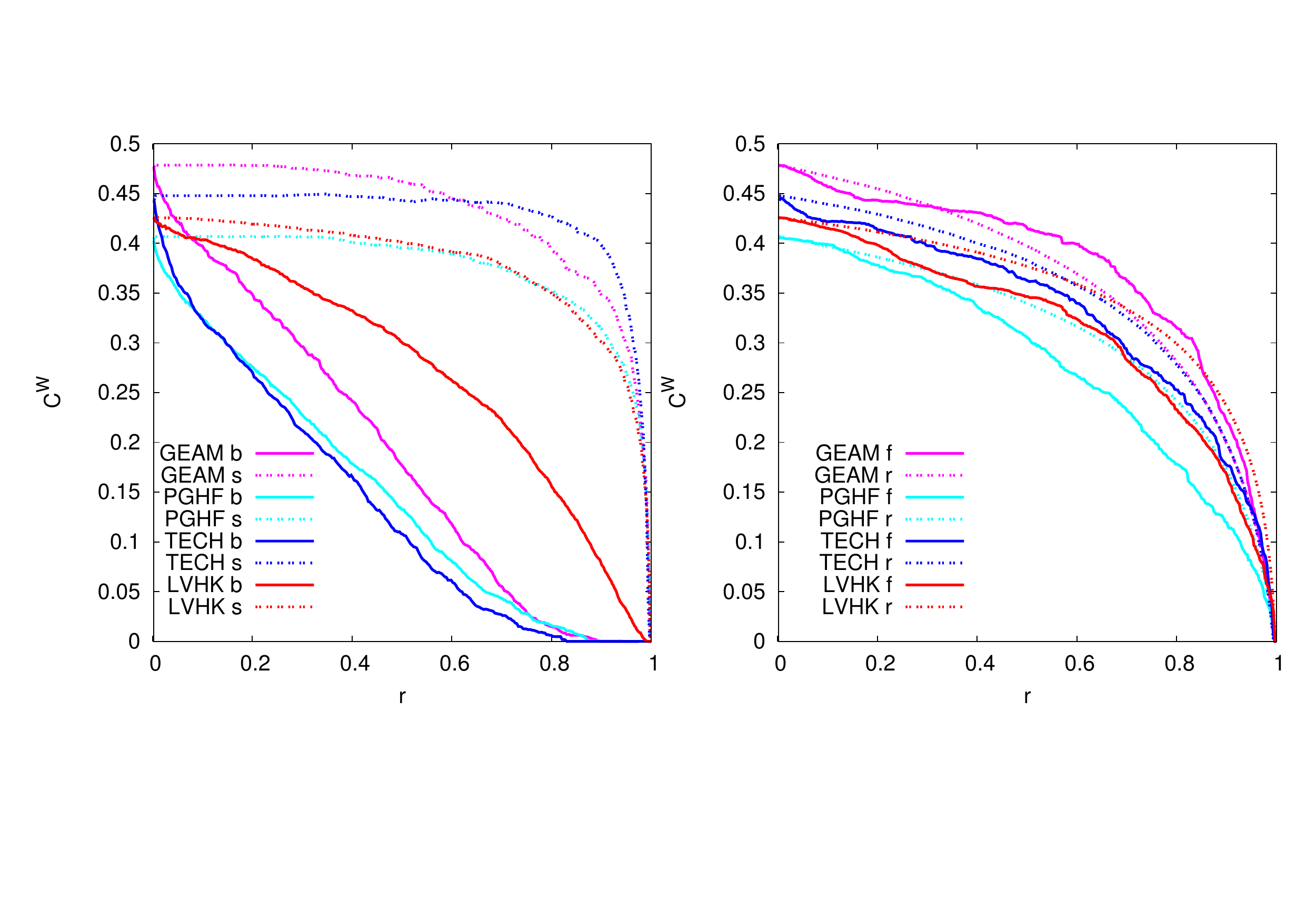}
\caption{\textbf{Importance of event size for the network cohesiveness.} Change of local network cohesiveness with removal of events according to their size (left) and temporal and random order (right). Abbreviations indicate order in which we remove events: {\bf b} - from the largest to the smallest, {\bf s} - from the smallest to the largest, {\bf f} - from the first to the last and {\bf r} - random.}
\label{Cw_event}
\end{figure}

Similar conclusions can be drawn based on the change of average weighted clustering coefficient $\langle C^{W} \rangle$ (now averaged over all nodes in the network) with the removal of events, Fig \ref{Cw_event}. Removal of events according to decreasing order of their sizes, does not result in the significant change of $\langle C^{W} \rangle$. The same value of weighted clustering coefficient, observed even after the removal of $80\%$ of events, shows that small events are not attended by \textit{a pair of} but rather \textit{by a group} of old friends. On the other hand, the removal of events in the opposite order results in gradual decrease of $\langle C^{W} \rangle$. A certain fraction of triads in networks are made by at least one link of low weight. These links are most likely to vanish after the removal of the largest events, which results in the gradual decrease of $\langle C^{W} \rangle$. Removal of events according to their temporal order results in the change of $\langle C^{W} \rangle$ similar to one obtained for random removal of events, confirming further that the time ordering of events does not influence the structure of studied networks.\\ 

\section*{Discussion and conclusion}
In this article we explore the event participation dynamics and underlying social mechanism of the Meetup groups. The motivation behind this was to further explore the event driven dynamics, work we have started by exploring participation patterns of scientists at scientific conferences \cite{smiljanic2016}, and to better examine the social origins behind the repeated attendance at group events, which was not feasible with the conference data. The results in this manuscript are based on empirical analysis of participation patterns and topological characteristics of networks for four different Meetup groups made up of people who have different motives and readiness to participate in group activities: GEAM, PGHF, TECH, LVHK.\\ 
Although these four groups differ in category and type of activity, we have shown that they are all characterized with similar participation patterns: the probability distributions of total number of participations, number of successive participations and time lag between two successive participations follow a power law and truncated power law behavior, with the value of power law exponents between $1$ and $3$. The resemblance of these patterns to those observed for conference participations \cite{smiljanic2016} indicates that these two, seemingly different, social system dynamics are governed by similar mechanism. This means that the probability for a member to participate in future events depends non-linearly on the balance between the numbers of previous participations and non-participations. As in the case of conferences \cite{smiljanic2016}, this behavior is independent of the group category, size, or location, meaning that members association with the community of a Meetup group strongly influence their event participation patterns, and thus the frequency and longevity of their engagement in the group activities.\\   
The Member's association with the community is primarily manifested trough her interconnectedness with other members of a specific Meetup group, i.e., in the structure of her personal social network. We have examined topological properties of filtered weighted social networks constructed from the members event co-occurrence. Trough network filtering we have emphasized the importance of significant links, the ones which are not the result of coincidence but rather an indicator of social relations. The analysis of local topological properties of these networks has revealed that the strength of connectedness with the community, for the members with small number of participations, is predominantly the consequence of the width of their social circles. Average strength and degree of members with $q\lesssim 50$, which on average corresponds to only a few participations, are equal, while the strength of members who know more than $50$ people and have participated in more than a few events, is several times higher then their degree. This means that after a few participations strengthening of existing ties becomes more important than meeting new people. These arguments are further extended with our observation of the evolution of average strength and degree with the growth of number of participations. Both, average degree and strength, grow, but the growth rate of strength is higher than one of the degree, for all four Meetup groups. All four groups are characterised with very high cohesiveness of their social communities. The evolutions of clustering coefficients, non and weighted one, and their ratio, show that bonding with the community becomes more important as the members' engagement in the group activity progresses. As in the case of conference participations, frequent attendees of group activities tend to form a core whose stability grows with the number of participations \cite{smiljanic2016,vandijk2006}. The need of frequent attendees to maintain and increase their bonding with the rest of the community influences their probability to attend future meetings and thus governs the event participation dynamics of the Meetup groups.\\
The observed structure of personal social networks of the Meetup members is in accordance with previous research on this topic \cite{dunbar1993, dunbar2008, hill2003, dunbar2011}. The average size of personal social networks for the most frequent attendees of the Meetup groups GEAM, PGHF, and TECH, is $150$ or lower, while the size of the LVHK personal network is less than $500$ different connections, i.e., of the same order. This is consistent with the predictions of the Social Brain Hypothesis for the typical human group size. The faster growth of the strength, compared to the one observed for degree, and the constant, non-trivial, value of the clustering coefficients are indicators of the layered structure of social networks. The comparable values of strength and degree, as well as weighted and non-weighted clustering coefficients, observed for small numbers of attendances, indicate that at the beginning all social connections are of the equal importance. As members' engagement with the community grows, she begins to interact with a certain members of the group more often, which results in the non-linear growth of her strength. The higher value of weighted clustering coefficient, compared to its non-weighted counterpart, indicates that member's personal network consists of layers, subgroups of members, characterized with similar strength of mutual relations.\\         
While the group category, type of activity and size do not significantly affect the participation dynamics in the group activities and structure of networks, the size of separate events does have an influence on the evolution of social networks. Large events represent an opportunity for members to make new acquaintances, i.e., to establish new connections. On the other hand, small meetings are typically the gatherings of members with preexisting connections, and their main purpose is to facilitate the stronger bonding among group members. We find that the time order of events is irrelevant for group dynamics.\\
The universality of the event participation patterns, shown in this and previous work \cite{smiljanic2016}, and its socially driven nature give us a better insight not only about the dynamics of studied social communities but also about others which are organised on very similar principles: communities that bring together people with the similar interests and where the participation is voluntary. Having in mind that these type of groups constitute a large part of human life, including all life aspects, understanding their functioning and dynamics is of great importance. Our results not only contribute to the corpus of increasing knowledge, but also indicate the key factor which influences the group longevity and successful functioning: the association of group members with the community. This and recent success stories \cite{websumit} suggest that complex network theory can be an extremely useful tool in creating successful communities. Future studies will be conducted towards further confirmation of universality of event participation patterns and better understanding of how social association and contacts can be used for creating conditions for successful functioning of learning and health support groups.
 
\section*{Materials and Methods}
\subsection*{Data}
There are more than $240000$ groups in $181$ countries classified into $33$ categories active in the Meetup community \cite{meetup}. For each of selected four groups, we have used the Meetup public API to access the data and collect the list of events organized by the group and the information on the members who confirmed their participation (RSVP) in the given event since the group's beginnings. Each member has a unique id which enables us to follow her activity in the group events during the time. The collected data have been fully anonymised and we did not collect any personal information about the group members. We have complied with terms of use of Meetup website. More details about the group sizes and the number of events is given in Table~\ref{dataset}.  
\begin{table}[h!]
\centering
\bigskip
\texttt{
  \begin{tabular}{|c|c|c|c|c|}
      \hline
  {\bf Meetup group} & {\bf Acronym} & {\bf Category} & $N_{m}$ &  $N_{e}$\\  \hline \hline
  geamcIt & GEAM & Food \& Drink & 5377 & 3986 \\ \hline
  pittsburgh-free& PGHF & Socializing & 4995 & 4617 \\ \hline
  techlifecolumbus & TECH & Tech & 3217 & 3162\\ \hline
  VegasHikers & LVHK & Outdoors \& Adventure & 6061 & 5096 \\ \hline
  \end{tabular}
}
\bigskip
\caption{Summary of collected data for four selected Meetup groups. $N_{m}$ is total number of group members, $N_{e}$ is total number of organised events.}
\label{dataset}
\end{table}

\subsection*{Network construction and filtering}
{\bf Network construction} We start with a bipartite member event network, which we represent with participation matrix $B$. Let $N_{m}$ denotes total number of members in the group and $N_{e}$ is total number of events organized by the group. If the member $i$ participated in the event $l$ element of matrix $B_{il}$ takes a value $1$, otherwise $B_{il}=0$. In the bipartite network created in this way, members' degree is equal to total number of events member participated in, while events' degree is defined as total number of members that have attended that event. The social network, which is the result of members interactions during the Meetup events and is represented by weighted matrix $W$, is created from the weighted projection of bipartite network to members partition \cite{mitrovic2010a,mitrovic2010b}. In the obtained weighted network nodes correspond to individual members while the value of the element of weighted matrix $W_{ij}$ corresponds to number of common events two members have attended together.

{\bf Network filtering} The observed weighted network is dense network where some of the non-zero edges can be the result of coincidence. For instance, these edges can be found between members who attended
large number of events or events with many participants, and therefore they do not necessarily indicate social connections between members. The pruning of these type of networks and separation of significant edges from non-significant ones is not a trivial task \cite{PhysRevE.93.012304, Dianati:2197182, 2016arXiv160702481S}. For this reasons we start from bipartite network and use method that determines the significance of $W_{ij}$ link based on configuration model of random bipartite networks \cite{saracco2015randomizing,2016arXiv160702481S,PhysRevE.93.032302,Dianati:2197182}. In this model of random networks the event size and the number of events a member attended are fixed, while all other correlations are destroyed (see SI for further explanations). Based on this model, for each link in bipartite network, $B_{il}$, we determine the probability $p_{il}$ that user $i$ has attended event $l$. The assumption of uncorrelated network enables us to also estimate the probability that two members, $i$ and $j$, have attended the same event, which is equal to $p_{il}p_{jl}$. Probability that two members have attended the same $w$ events is then given by Poisson binomial distribution 
\begin{equation}
 P_{ij}(w) = \sum_{M_w} \prod_{l \in M_w}p_{il}p_{jl} \prod_{\bar{l} \notin M_w}(1-p_{i\bar{l}}p_{j\bar{l}}) \,
\end{equation}
where $M_w$ is the subset of $w$ events that can be chosen from given $M$ events \cite{0295-5075-113-2-28003, 2016arXiv160702481S, Dianati:2197182}. We define $p$-value as probability that two members $i$ and $j$ has co-occurred on at least $w_{ij}$ events, i.e., that the link weight between these two members is $w_{ij}$ or higher
\begin{equation}
 p\text{-value}(w_{ij})=\sum_{w \geq w_{ij}}P_{ij}(w).
\end{equation}
The relationship between users $i$ and $j$ will be considered statistically significant if $p\text{-value}(w_{ij}) \leq p_{trs}$. In our case, threshold $p_{trs}=0.05$. All links with $p\text{-value}(w_{ij}) > p_{trs}$ are consequence of chance and are considered as non-significant and thus removed from the network. This way we obtain weighted social network of significant relations between members of the Meetup group $W^{S}_{ij}$. The details on how we estimate $p_{il}$ and $P_{ij}(w)$ for each link are given in SI.\\
\textbf{Topological measures} All topological measures considered in this work are calculated for weighted social network of significant relations $W^{S}_{ij}$. We consider the following topological measures of the nodes:
\begin{itemize}
\item The node degree $q_{i}=\sum_{j}\mathcal{H}(W^{S}_{ij})$, where $\mathcal{H}$ is Heaviside function ($\mathcal{H}(x)=1$ if $x>0$ otherwise $\mathcal{H}(x)=0$); 
\item The node strength $s_{i}=\sum_{j}W^{S}_{ij}$ \cite{boccaletti2006};
\item Non-weighted clustering coefficient of the node $c_{i}=\frac{1}{q_{i}(q_{i}-1)}\sum_{j,m}\mathcal{H}(W^{S}_{ij})\mathcal{H}(W^{S}_{im})\mathcal{H}(W^{S}_{jm})$ \cite{boccaletti2006}.
\item Weighted clustering coefficient of the node $c^W_i = \frac{1}{s_i(q_i - 1)}\sum_{jm}\frac{W_{ij}^{S}+W_{im}^{S}}{2}\mathcal{H}(W_{ij}^{S})\mathcal{H}(W_{im}^{S})\mathcal{H}(W_{jm}^{S})$ \cite{Barrat16032004}.
\end{itemize}
Weighted clustering coefficient of the network $\langle C^{W} \rangle$ and its non-weighted counterpart $\langle C \rangle$ are values averaged over all nodes in the network.

\subsection*{The event relevance}
In order to explore the relevance of event size and time ordering for the evolution of social network topology we analyze how removal of events, according to specific ordering, influences the number of acquaintance and network cohesion. Specifically, we observe change of measure $\eta$, which represents the fraction of the remaining acquaintances, and weighted clustering coefficient $\langle C^{W} \rangle$ after the removal of a fraction $r$ of events. The removal of event results in change of link weights between group members. For instance, if two members, $i$ and $j$, have participated in event $l$, the removal of this event will result in the decrease of the link weight $W^{S}_{ij}$ by one. Further removal of events in which these two members have co-occurred will eventually lead to termination of their social connection, i.e., $W^{S}_{ij}=0$. If $W^{S}(r)$ is the matrix of link weights after the removal of a fraction $r$ of events and $W^{S}$ is the original matrix of significant relations, then the value of parameter $\eta$ after the removal of $r$ events is calculated as 
\begin{equation}
 \eta(r) = \frac{\sum_{ij}\mathcal{H}(W^{S}_{ij}(r))}{\sum_{ij}\mathcal{H}(W^{S}_{ij})} \ ,
\end{equation}  
The value of weighted clustering coefficient $\langle C^{W} \rangle$ after the removal of a fraction $r$ of events is calculated using the same formula as for the $\langle C^{W} \rangle$ just using the value of $W^{S}(r)$ instead of $W^S$.\\
We remove events according to several different strategies:
\begin{itemize}
 \item We sort events according to their size. Then, we remove sorted events in descending and ascending order. 
 \item We remove events according to their time-order, from the first to the last.
 \item We remove events in random order. We perform this procedure for each list of events $100$ times.
\end{itemize}

\section*{Acknowledgments}
Numerical simulations were run on the PARADOX supercomputing facility at the Scientific Computing Laboratory of the Institute of Physics Belgrade.

\section*{Supporting Information}

\renewcommand{\thefigure}{S\arabic{figure}}
\setcounter{figure}{0}

\renewcommand{\thetable}{S\arabic{table}}
\setcounter{table}{0}

\subsection{Network filtering}

The Meetup dataset, containing information on organised events by certain Meetup group and members of that group that confirmed attendance at an event, allows us to construct member-event bipartite network with adjacency matrix $B$. For each member $i\in\{1,\dots, N\}$ and event $l\in\{1,\dots,M\}$, matrix element $B_{il}=1$ if member $i$ participated event $l$, 
or $B_{il}=0$, otherwise. The degree of member $i$ is defined as the total number of events member $i$ participated in, $k_i=\sum_{l}B_{il}$, and similarly, the degree of event $l$
is defined as the total number of members attended the event, $d_l=\sum_{i}B_{il}$. Given the matrix $B$, social relations between Meetup members can be analysed using the projected unipartite member-member weighted network, where the weight of the link between two members is equal to the number of events they both attended. The observed weighted network is the dense network where some of the non-zero edges can be a matter of coincidence. For instance, two frequent attendees can meet several times due to a chance not due to the fact that there is some relation between them, which means that the connection between them is not significant for our analysis. Also, the connections between members that meet at big events and never again can not be regarded as social relations and thus they need to be excluded from our analysis. To make the distinction between significant and non-significant edges is nontrivial task \cite{PhysRevE.93.012304, Dianati:2197182, 2016arXiv160702481S}. Here we use the method which enables us to calculate the significance of the link between two members based on the probability for that link to occur in random network. As a null model we use configuration model of bipartite network \cite{saracco2015randomizing,2016arXiv160702481S,PhysRevE.93.032302,Dianati:2197182}.

First we describe general framework for constructing randomized network ensemble $\mathcal{G}$ with given structural constraints $\{x_i\}$. The maximum-entropy probability of the 
graph in the ensemble, $P(G)$, is given by
\begin{equation}
 P(G) = \frac{1}{Z}e^{-\sum_{i}\lambda_{i}x_{i}},
\end{equation}
where the $\lambda_{i}$ are Lagrangian multipliers and the partition function of these network ensembles is defined as
\begin{equation}
 Z = \sum_{G}e^{-\sum_{i}\lambda_{i}x_{i}}. \label{Z}
\end{equation}
The ensemble average of a graph property $x_{i}$ can be expressed as
\begin{equation}
 \langle x_i \rangle = \sum_{G} x_{i}(G) P(G) = -\frac{\partial}{\partial \lambda_{i}} \ln Z. \label{constraints}
\end{equation}
Then the constants $\lambda_i$ could be determined from (\ref{constraints}).

Let us now consider configuration model of the member-event bipartite network with given degree sequence $k_i$ and $d_l$. In this case the partition function can be written as
\begin{equation}
 Z = {\sum_{G}e^{-\sum_{i}\alpha_{i}k_{i}-\sum_{l}\beta_{l}d_{l}}} = \sum_{G}e^{-\sum_{il}(\alpha_{i}+\beta_{l})B_{il}} = \prod_{il}(1+e^{-(\alpha_{i}+\beta_{l})}).
\end{equation}
The Lagrangian multipliers $\alpha_i$ and $\beta_l$ are determined from
\begin{equation}
 k_i = -\frac{\partial}{\partial \alpha_{i}} \ln Z = \sum_{l=1}^{M}\frac{e^{-\alpha_{i}-\beta_{l}}}{1 + e^{-\alpha_{i} - \beta_{l}}},
\end{equation}
\begin{equation}
 d_l = -\frac{\partial}{\partial \beta_{l}} \ln Z = \sum_{i=1}^{N}\frac{e^{-\alpha_{i}-\beta_{l}}}{1 + e^{-\alpha_{i} - \beta_{l}}}.
\end{equation}
Finally, we can calculate the probability $p_{il}$ that a member $i$ attended event $l$. If we define coupling parameter $\lambda_{il}=\alpha_{i}+\beta_{l}$ and write partition
function in the form
\begin{equation}
 Z = \sum_{G}e^{-\sum_{il}\lambda_{il}B_{il}} = \prod_{il}(1+e^{-\lambda_{il}}),
\end{equation}
then, it holds
\begin{equation}
 p_{il} = \langle B_{il} \rangle = -\frac{\partial}{\partial \lambda_{il}} \ln Z = \frac{e^{-\lambda_{il}}}{1+e^{-\lambda_{il}}} = \frac{e^{-\alpha_{i}-\beta_{l}}}{1 + e^{-\alpha_{i} - \beta_{l}}}.
\end{equation}

Now, when the probability $p_{il}$ is given, the probability that members $i$ and $j$ both participated in event $l$ is $p_{ij}(l) = p_{il}p_{jl}$. The probability $P_{ij}(w)$ 
of having an edge of the weight $w$ between the nodes $i$ and $j$ is given by Poisson binomial distribution
\begin{equation}
 P_{ij}(w) = \sum_{M_w} \prod_{l \in M_w}p_{ij}(l) \prod_{\bar{l} \notin M_w}(1-p_{ij}(\bar{l})),
\end{equation}
where $M_w$ is the subset of $w$ events that can be chosen from given $M$ events \cite{0295-5075-113-2-28003, 2016arXiv160702481S, Dianati:2197182}. We use DFT-CF method (Discrete
Fourier Transform of characteristic function), proposed in \cite{Hong201341}, to compute Poisson binomial distribution.

On the basis of $P_{ij}(w)$, we define $p$-value as the probability that edge $(i,j)$ has weight higher or equal than $w_{ij}$
\begin{equation}
 p\text{-value}(w_{ij})=\sum_{w \geq w_{ij}}P_{ij}(w).
\end{equation}
The edge $(i,j)$ will be considered statistically significant if $p\text{-value}(w_{ij}) \leq \alpha$. In our case, threshold $\alpha=0.05$. If $p\text{-value}(w_{ij}) > \alpha$, 
the edge $(i,j)$ should be removed as spurious statistical connection between members (set $w_{ij}=0$).

\subsection{Distribution fitting}
\begin{figure}
\begin{center}
  \includegraphics[scale=0.85]{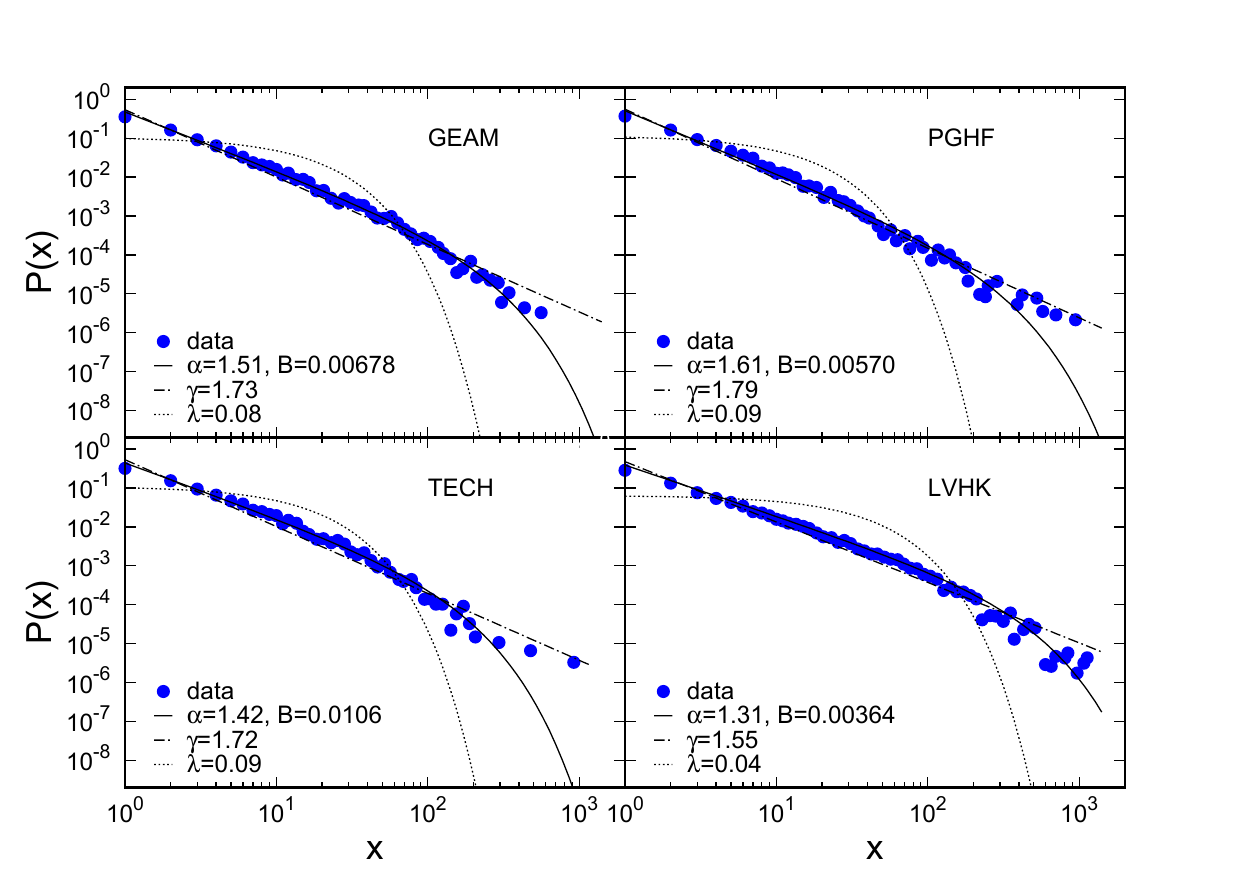}
 \end{center}
\caption{The probability distribution $P(x)$ of total numbers of participations in group events $x$, obtained from the empirical data for the four selected Meetup groups (blue circles). We also show truncated power law fit $x^{-\alpha}e^{-Bx}$ (solid lines), power law fit $x^{-\gamma}$ (dotted-dashed lines), and exponential fit $e^{-\lambda x}$ (dotted lines).} 
\label{fig_A}
\end{figure}

We fit exponential function $e^{-\lambda x}$, power law function $x^{-\gamma}$ and truncated power law $x^{-\alpha}e^{-Bx}$ to the probability distribution of the total number
of participations in group events using the maximum-likelihood fitting method~\cite{doi:10.1137/070710111}. It is evident from Fig~\ref{fig_A} that the distribution does not follow 
exponential fit. We compare how the power law and the truncated power law distribution, which are the nested versions of each other, fit the data by calculating the log likelihood ratio $\mathcal{R}$ and $\pi$-value (see Ref.~\cite{doi:10.1137/070710111}). Here, the negative value of $\mathcal{R}$ indicates that the truncated power law is a superior fit to the 
power law. Additionally, when the value of $\mathcal{R}$ tends to $0$, one can use $\pi$-value. The small $\pi$-value indicates that the power law distribution can be excluded. 
Table~\ref{table_A} shows that the truncated power law is a superior fit compared to power law for all four empirical distributions.

\begin{table}
 \begin{center}
  \begin{tabular}{|C{2.5cm}|C{2.5cm}|C{2cm}|}
      \hline
                    & $\mathcal{R}$ & $\pi$     \\  \hline \hline
      GEAM & -97.70 & 0.0  \\ \hline
      PGHF & -58.04 & 0.0  \\ \hline
      TECH & -90.84 & 0.0  \\ \hline
      LVHK & -236.52 & 0.0 \\ \hline
  \end{tabular}
 \end{center}
\caption{Log likelihood ratio $\mathcal{R}$ and the $\pi$-value compare fits to the power law and fits to the truncated power law for the probability distribution of total numbers of participations in group events.}
\label{table_A}
\end{table}

\subsection{Data randomization}
We randomize event participation patterns preserving the total number of participation for each member and the number of participants per event \cite{strona2014fast}.
Firstly, we choose at random two members, $i$ and $j$, and for each of them we choose randomly an event they participated, $l^i$ and $l^j$. If $l^i \neq l^j$, and $i$ didn't 
participate at event $l^j$ and $j$ didn't participate at event $l^i$, they are swapped. We perform 10 $\times$ (number of participants) $\times$ (number of events) swaps.
The randomization of the event participation times induces transformations of associated weighted network. The number of participants at some event will stay the same, but participants will differ, resulting in weight increase of certain edges and likewise in weight decrease of some other edges in weighted network. The total weight of the network will be preserved. 

For each Meetup group we generate 100 randomized weighted networks and filter out non-significant edges. 

\subsection{Figures}

\begin{figure}
\begin{center}
  \includegraphics[scale=0.85]{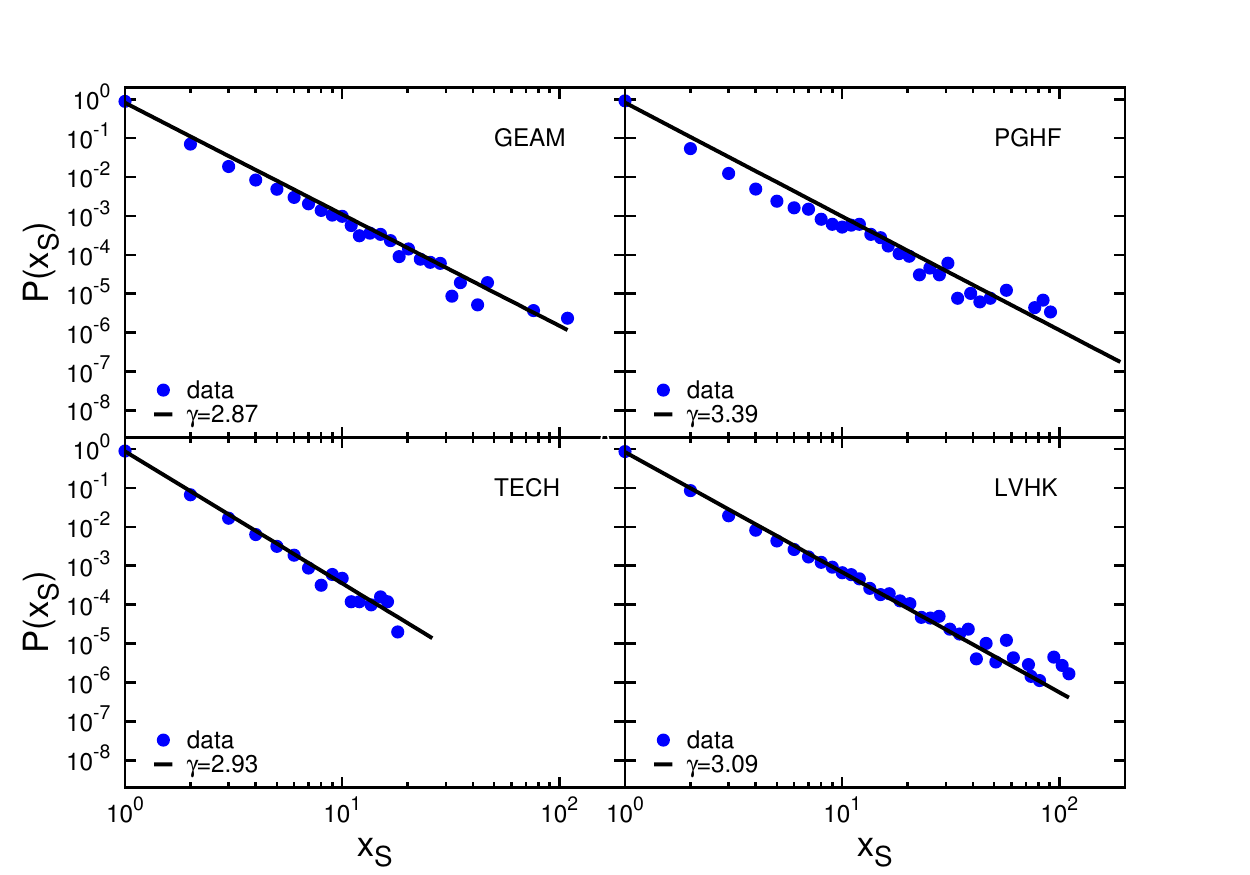}
 \end{center}
\caption{The probability distribution of successive numbers of participations in group events $x_{S}$, for the four selected Meetup groups. The probability distribution follows power law behavior $P(x_{S})\sim x_{S}^{-\gamma}$.} 
\label{fig_B}
\end{figure}

\begin{figure}
\begin{center}
  \includegraphics[scale=0.85]{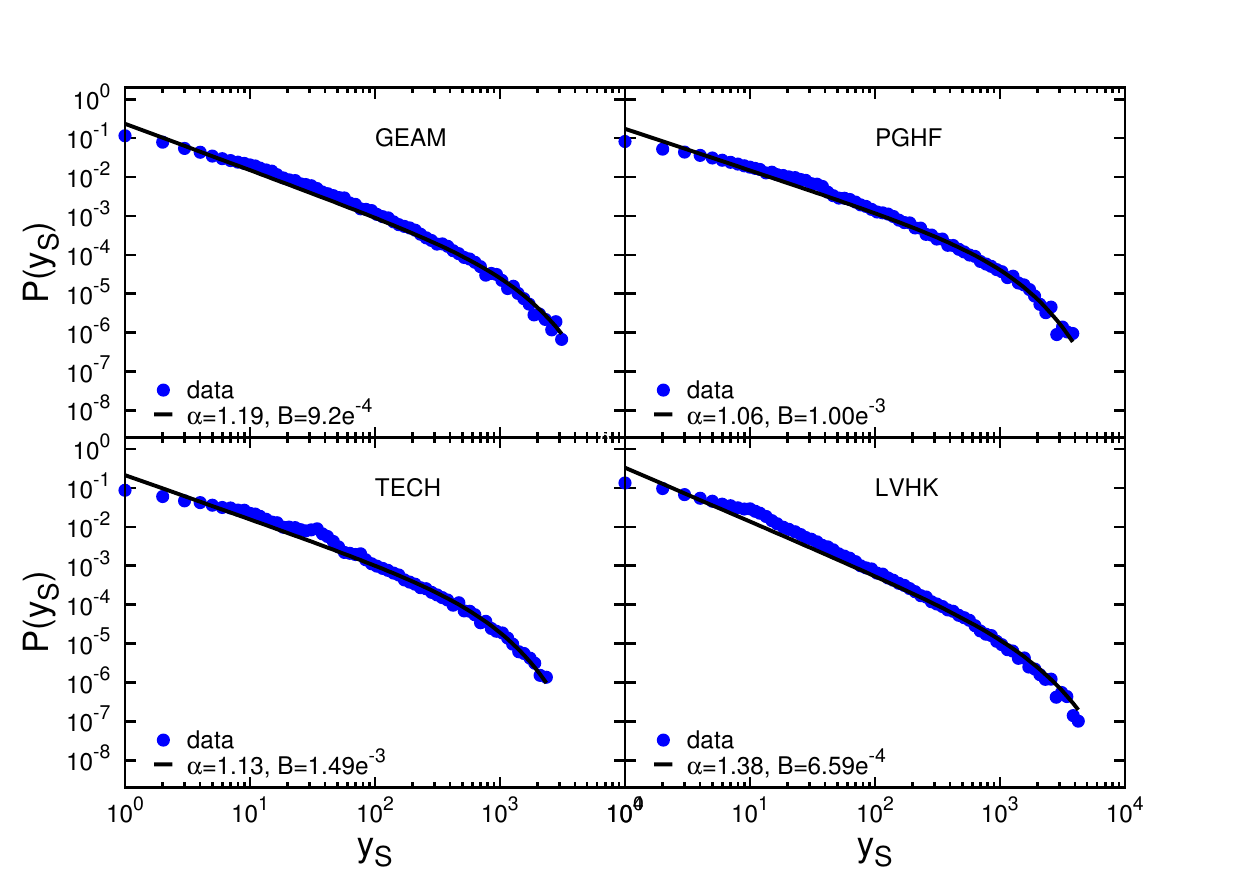}
 \end{center}
\caption{The probability distribution of time lags between two successive participations in group events $y_{S}$, for the four selected Meetup groups. The probability distribution follows truncated power law behavior $P(y_{S})\sim y_{S}^{-\alpha}e^{-B y_{S}}$.} 
\label{fig_C}
\end{figure}

\begin{figure}
\begin{center}
  \includegraphics[scale=0.85]{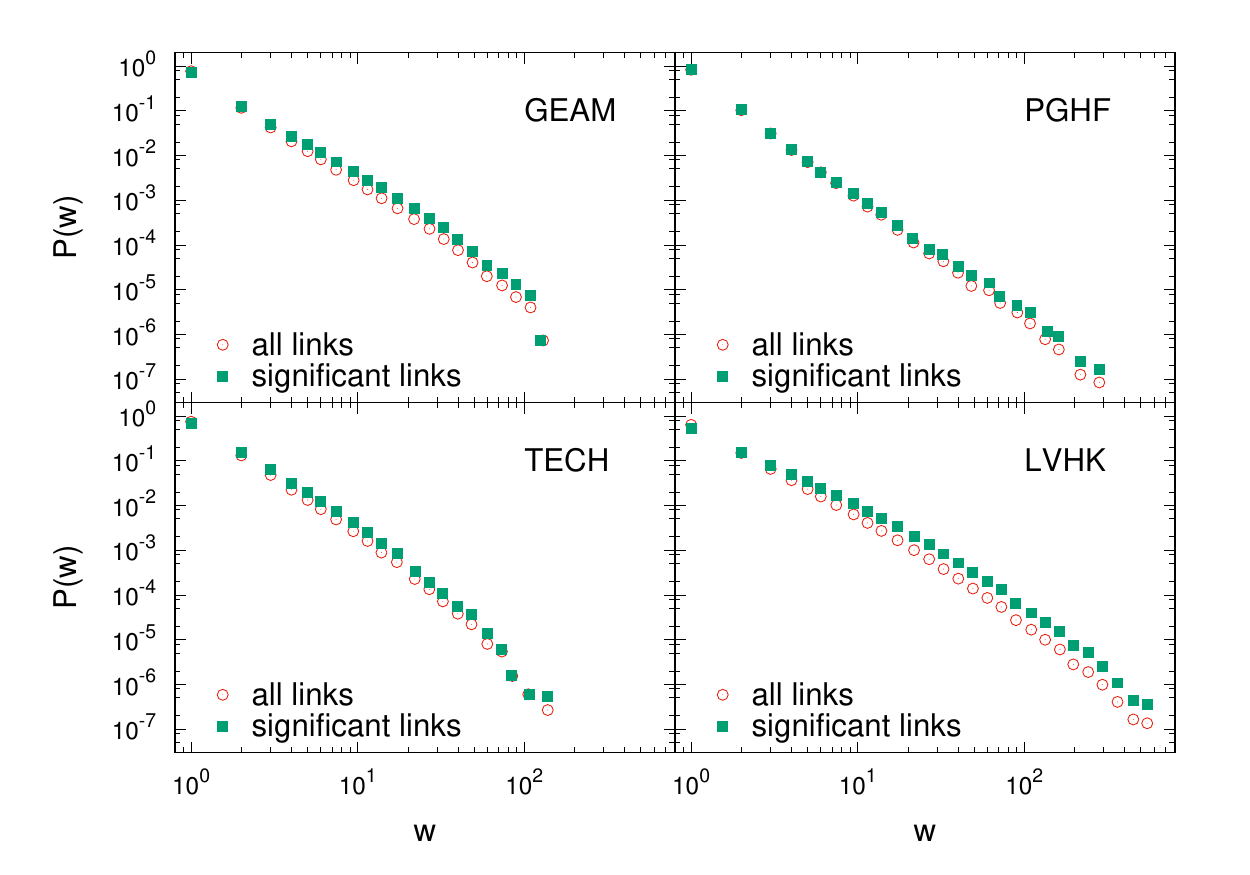}
 \end{center}
\caption{The probability distribution of link weights in a weighted network before and after filtering, for the four selected Meetup groups.} 
\label{fig_D}
\end{figure}

\begin{figure}
\begin{center}
  \includegraphics[scale=0.85]{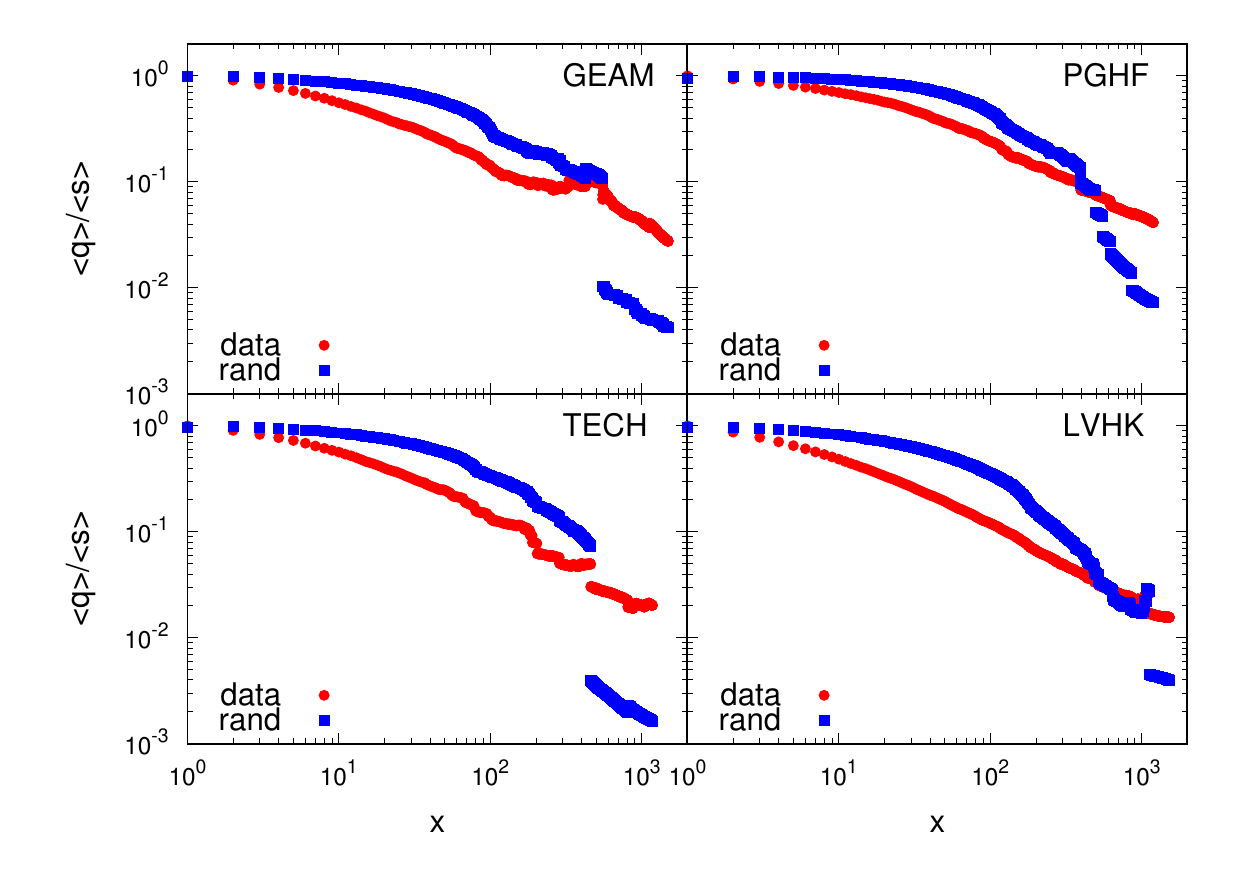}
 \end{center}
\caption{The dependence of a degree strength ratio on the number of participations, averaged over all members for the four considered Meetup groups. Red circles correspond to results obtained from empirical data, while blue squares correspond to randomized data.} 
\label{fig_E}
\end{figure}

\begin{figure}
\begin{center}
  \includegraphics[scale=0.85]{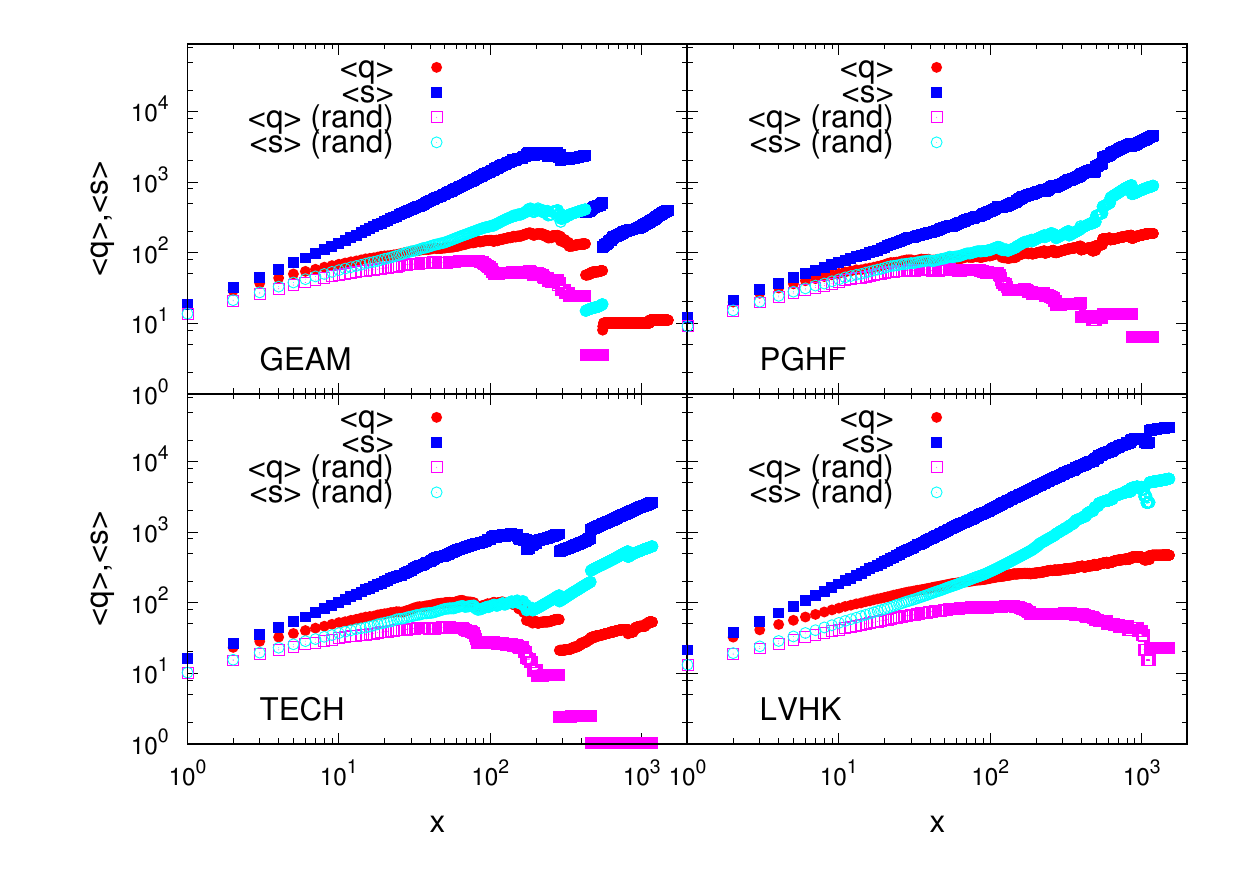}
 \end{center}
\caption{The dependence of group members' average degree $\langle q \rangle$ and strength $\langle s \rangle$ on numbers of participations for a real weighted network and a randomized network.} 
\label{fig_F}
\end{figure}

\begin{figure}
\begin{center}
  \includegraphics[scale=0.85]{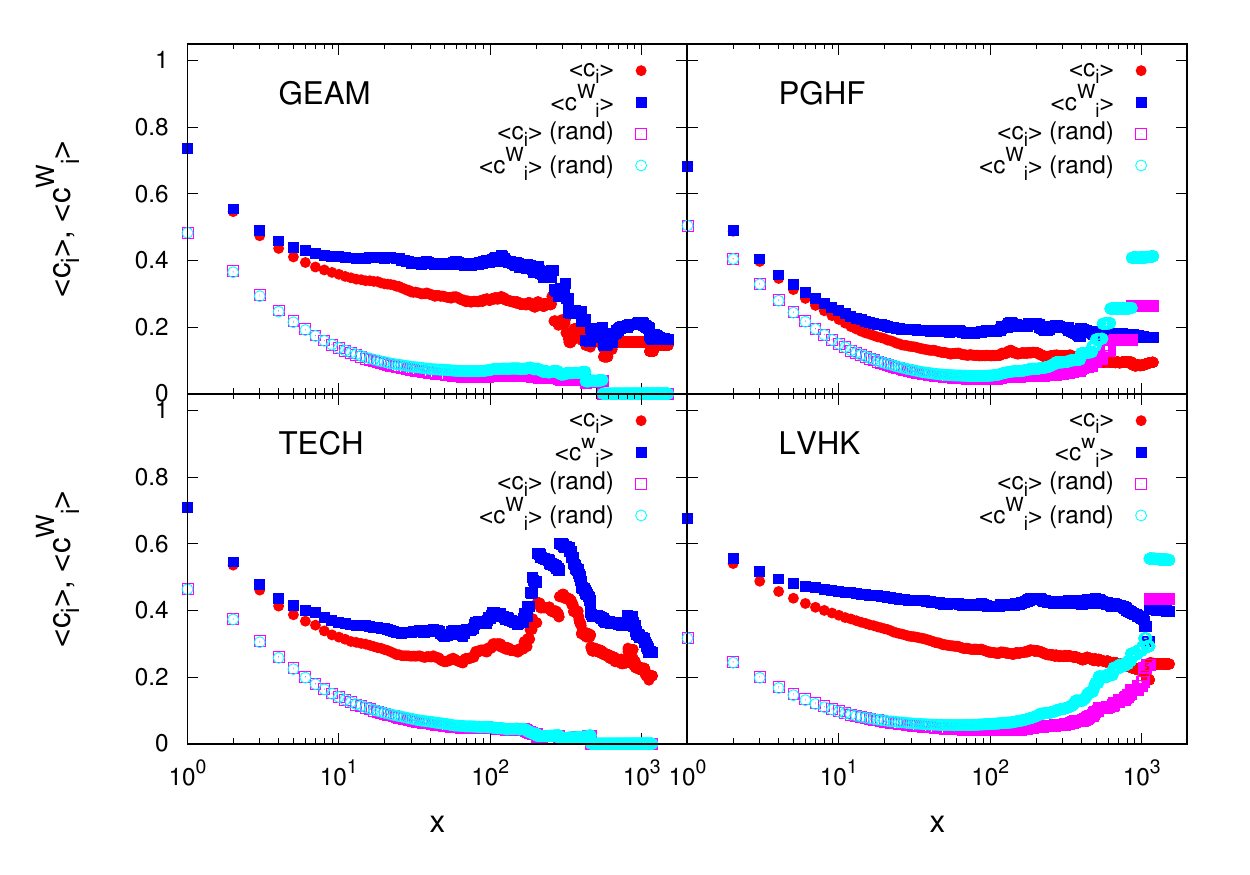}
 \end{center}
\caption{The dependence of group members' average non-weighted $\langle c_i \rangle$ and weighted clustering coefficient $\langle c_i^W \rangle$ on numbers of participations for a real weighted network and a randomized network.} 
\label{fig_G}
\end{figure}

\begin{figure}
\begin{center}
  \includegraphics[scale=0.85]{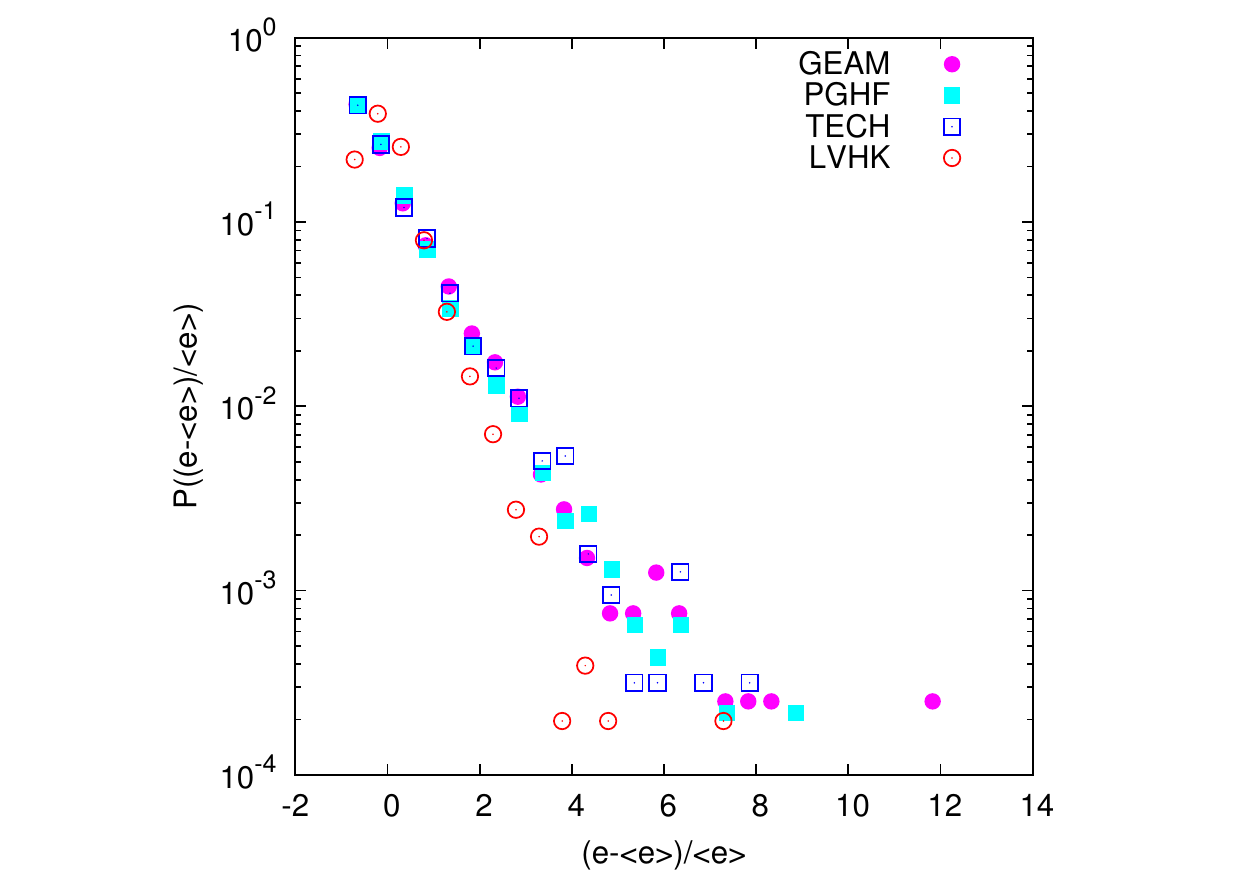}
 \end{center}
\caption{The probability distribution of relative size fluctuations $\frac{\langle e \rangle - e}{\langle e \rangle}$, for the four considered Meetup groups, where $e$ is the event size and $\langle e \rangle$ is the average event size.} 
\label{fig_H}
\end{figure}

\end{document}